\def\spose#1{\hbox to 0pt{#1\hss}}
\def\approxlt{\mathrel{\spose{\lower 3pt\hbox{$\sim$}}
        \raise 2.0pt\hbox{$<$}}}
\def\approxgt{\mathrel{\spose{\lower 3pt\hbox{$\sim$}}
        \raise 2.0pt\hbox{$>$}}}

\def\multleft#1{\hbox to size{\vbox {\halign {\lft{##}\cr #1}}\hfill}\par}
\def\multright#1{\hbox to size{\vbox {\halign {\rt{##}\cr #1}}\hfill}\par}

\def\degmark{^\circ}
\def\s{\hbox{\phantom{5}}}      

\def\boxit#1{\vbox{\hrule\hbox{\vrule\kern3pt\vbox{\kern3pt
          #1 \kern3pt}\kern3pt\vrule}\hrule}}

\def\cm{{\rm\thinspace cm}}

\def\erg{{\rm\thinspace erg}}
\def\eV{{\rm\thinspace eV}}

\def\keV{{\rm\thinspace keV}}

\def\Msun{\hbox{$\rm\thinspace M_{\odot}$}}

\def\s{{\rm\thinspace s}}

\def\ergps{\hbox{$\erg\s^{-1}\,$}}

\def\ps{\hbox{$\s^{-1}\,$}}

\documentclass[12pt]{article}

\usepackage{psfig}
\usepackage{aasms4}
\usepackage{times}
\usepackage{amssym}

\makeatletter
\@ifundefined{chapter}{\def\thebibliography#1{
\section*{References}
  \list
  {\relax}{\setlength{\labelsep}{0em}
        \setlength{\itemindent}{-\bibhang}
        \setlength{\itemsep}{\parskip}
        \setlength{\parsep}{0pt}
        \setlength{\leftmargin}{\bibhang}}
    \def\newblock{\hskip .11em plus .33em minus .07em}
    \sloppy\clubpenalty4000\widowpenalty4000
    \sfcode`\.=1000\relax}}%
{\def\thebibliography#1{
  \list
  {\relax}{\setlength{\labelsep}{0em}
        \setlength{\itemindent}{-\bibhang}
        \setlength{\itemsep}{\parskip}
        \setlength{\parsep}{0pt}
        \setlength{\leftmargin}{\bibhang}}
    \def\newblock{\hskip .11em plus .33em minus .07em}
    \sloppy\clubpenalty4000\widowpenalty4000
    \sfcode`\.=1000\relax}}
 
\newlength{\bibhang}
\let\@internalcite\cite
\def\cite{\@ifstar{\citey}{\citefull}}
\def\citefull{\def\astroncite##1##2{##1\ ##2}\@internalcite}
\def\citey{\def\astroncite##1##2{##1\ (##2)}\@internalcite}
\def\citeyear{\def\astroncite##1##2{##2}\@internalcite}
\def\citename{\def\astroncite##1##2{##1}\@internalcite}
\def\@citex[#1]#2{\if@filesw\immediate\write\@auxout{\string\citation{#2}}\fi
  \def\@citea{}\@cite{\@for\@citeb:=#2\do
    {\@citea\def\@citea{; }\@ifundefined
       {b@\@citeb}{{\bf ??}\@warning
       {Citation `\@citeb' on page \thepage \space undefined}}%
{\csname b@\@citeb\endcsname}}}{#1}}
 
\def\@cite#1#2{#1\if@tempswa #2\fi} 
\def\@biblabel#1{}
 
\def\astroncite#1#2{#1\ #2}
\makeatother

\begin{document}
 
\title{X-ray iron line reverberation from black hole accretion disks}

\author{Christopher~S.~Reynolds\altaffilmark{1},
   Andrew~J.~Young\altaffilmark{2}, Mitchell~C.~Begelman\altaffilmark{1,3},
   Andrew~C.~Fabian\altaffilmark{2}}
 
 \altaffiltext{1}{JILA, University of Colorado, Campus Box 440, Boulder, CO 80309-0440\\
   chris@rocinante.colorado.edu, mitch@jila.colorado.edu}

\altaffiltext{2}{Institute of Astronomy, Madingley Road, Cambridge CB2~OHA,
  UK.\\\{ayoung, acf\}@ast.cam.ac.uk}
 
\altaffiltext{3}{Department of Astrophysical and Planetary Sciences,
University of Colorado, Boulder, CO 80309-0391.}

\begin{abstract}
  The relativistically broad X-ray iron line seen in many AGN spectra is
  thought to originate from the central regions of the putative black hole
  accretion disk.  Both the line profile and strength will vary in response
  to rapid variability of the primary X-ray continuum source.  The temporal
  response of the line contains information on the accretion disk
  structure, the X-ray source geometry, and the spin of the black hole.
  Since the X-ray source will have a size comparable to the fluorescing
  region of the accretion disk, the general reverberation problem is not
  invertible.  However, progress can be made since, empirically, AGN light
  curves are seen to undergo dramatic short timescale variability which
  presumably corresponds to the creation of a single new active region
  within the distributed X-ray source.  The iron line response to these
  individual events can be described using linear transfer theory.
  
  We consider the line response to the activation/flaring of a new X-ray
  emitting region.  Most of our detailed calculations are performed for the
  case of an X-ray source on the symmetry axis and at some height above the
  disk plane around a Kerr black hole.  We also present preliminary
  calculations for off-axis flares.  We suggest ways in which future,
  high-throughput X-ray observatories such as {\it XMM} and the {\it
    Constellation X-ray Mission} may use these reverberation signatures to
  probe both the mass and spin of AGN black holes, as well as the X-ray
  source geometry.
\end{abstract}

\begin{keywords}
{accretion, accretion disks --- black hole physics --- galaxies: active ---
  galaxies: Seyfert --- X-rays: galaxies}
\end{keywords}

\section{Introduction}
 
Recent X-ray spectroscopy has provided the first direct observational probe
of the innermost regions of black hole accretion disks.  X-ray irradiation
of the surface layers of the inner accretion disk by a disk-corona results
in the emission of the prominent fluorescent K$\alpha$ line of iron in the
X-ray regime (George \& Fabian 1991; Matt, Perola \& Piro 1991).  The
combined action of mildly-relativistic Doppler shifts (due to the orbital
motion of the accretion disk material) and strong gravitational redshifts
broaden and skew this line in a dramatic and very characteristic manner
(Fabian et al.  1989; Laor 1991).  Spectroscopy of bright Seyfert nuclei
with the Advanced Satellite for Cosmology and Astrophysics ({\it ASCA})
confirms that such lines do indeed possess exactly this broadened/skewed
profile (Tanaka et al. 1995; Nandra et al.  1997a).  Moreover, alternative
broadening/skewing mechanisms (i.e., those not involving black hole
accretion disks) appear to fail (Fabian et al. 1995).  This is the first
direct evidence for a relativistic accretion disk in an AGN.  Furthermore,
this emission line provides us with a ``clock'' in orbit about a black hole
with which we can study strong gravity in detail.

The observed line profile depends upon the space-time geometry, the
accretion disk structure, and the pattern of X-ray illumination across the
surface of the disk.  It is well known that the X-ray source in many AGN is
highly variable on short timescales.  For example, during {\it ASCA}
performance verification observations of the bright Seyfert galaxy
MCG$-$6-30-15, the {\it ASCA} band X-ray flux was observed to undergo a
step-like doubling during a period of $\sim 100\s$ or less (Reynolds et al.
1995).  The timescale of this event is probably comparable to, or less
than, the light crossing time of the black hole event horizon (see Section
7).  Physically, this event most likely corresponds to the activation of a
new flaring region within the X-ray emitting disk-corona.  The subsequent
change in the X-ray illumination pattern on the accretion disk, coupled
with the finite light travel times within the system, will lead to temporal
changes in the iron line profile and strength.  Such changes are known as
reverberation effects.

Even when observing a bright Seyfert galaxy, {\it ASCA} (which to this date
represents the state-of-the-art in X-ray spectrometers) can only achieve a
count rate in the iron line of $10^{-2}\,{\rm photon}\,{\rm s}^{-1}$.  The
long integration times required to define the line strength and profile
given this count rate ($\sim 1$\,day) will average over the reverberation
effects described in the above paragraph.  While still providing extremely
exciting results, these {\it time-averaged} studies will always have
limitations.  First, the length scales relevant to the time-averaged lines
are expressible purely in terms of the gravitational radius $r_{\rm
  g}=Gm/c^2$, where $m$ is the mass of the central black hole.  Thus,
time-averaged line profiles alone cannot determine the absolute value of
$r_{\rm g}$ and hence the mass of the black hole.  Second, fitting
time-averaged models to time-averaged line profiles results in a degeneracy
which prevents one from disentangling the geometry of the X-ray source, the
structure of the accretion disk, and the spin of the black hole.  The
`very-broad' state of the iron line in MCG$-$6-30-15 as found by Iwasawa et
al. (1996) provides an excellent illustration of this problem.  By assuming
that the X-ray source is a thin corona over the surface of the disk with an
X-ray flux proportional to the viscous dissipation in the underlying disk
(Page \& Thorne 1974), the models of Dabrowski et al. (1997) suggest that
the black hole must be rapidly spinning with a dimensionless spin parameter
of $a>0.94$.  However, Reynolds \& Begelman (1997; hereafter RB97) present
an alternative scenario in which a high latitude X-ray source, displaced
somewhat from the disk, excites appreciable fluorescence from inside the
radius of marginal stability.  They show that the MCG$-$6-30-15 result is
compatible with a line emitted from Schwarzschild geometry (i.e., a
non-rotating black hole).  Young et al. (1998) demonstrated that a
significant Compton reflection component (primarily a strong iron edge) is
predicted by the RB97 scenerio.  However, it is a subject of debate whether
current data are able to rule out the RB97 picture via the non-detection of
the reflected continuum/edges.

Reverberation effects hold the key to unlocking the full diagnostic power
of broad iron lines.  Iron line reverberation was first explicitly
discussed in detail by Stella (1990).  He calculated the detailed line
variability from a disk around a Schwarzschild black hole given a
step-function doubling in the flux of the X-ray source which was assumed to
be at the exact geometric center of the disk.  This work was extended by
Matt \& Perola (1992) and Campana \& Stella (1993, 1995), who calculated
the temporal behavior of moments of the line (considered by these authors
to be more observationally convenient), as well as some alternative source
geometries.  The principal goal of these studies was to suggest how iron
line reverberation can be used to measure the mass of the black hole.

In this paper we complement and extend this previous work in the light of
current observational issues.  In particular, our model for computing iron
line reverberation assumes a black hole with an arbitrary spin parameter
(assumed to be consistent with the cosmic censorship hypothesis, i.e.,
$|a|<1$) and we search for diagnostics of this spin in the observed
reverberation signatures.  In Section~2 we discuss the general structure of
the iron line reverberation problem and show it to be qualitatively more
difficult than reverberation mapping of the broad line region (BLR; see
Peterson 1993 for a review of BLR reverberation mapping techniques).  In
Section~3 we outline our assumptions and discuss their immediate
consequences.  In Section~4 we discuss the computation of (time-dependent)
iron line profiles.  Detailed reverberation calculations for on-axis
sources are presented in Section~5.  In Section~6 we briefly address the
issue of off-axis flares and show two preliminary off-axis calculations.
To illustrate the use of these techniques, we discuss the determination of
the X-ray source geometry and black hole spin for the case of MCG$-$6-30-15
in Section~7.  Our conclusions are drawn together in Section~8.

\section{The general nature of the problem}

The techniques and formalism of reverberation have been applied extensively
to the problem of mapping the broad line region (BLR) of AGN.  The size of
the BLR (i.e., light weeks to light months) is very large compared to the
source of ionizing UV/X-ray photons.  Hence the problem that these studies
address is the following: what geometry of line-emitting plasma is required
to reproduce the observed line variability given excitation by a (variable)
point source of radiation at the center.  Assuming that the {\it local}
line emission is linearly proportional to the instantaneous ionizing flux,
one can write a linear transfer equation
\begin{equation}
F_{\rm l}(\nu,t)=\int_{-\infty}^{\infty}F_{\rm c}(t-t^\prime)\psi(\nu,t^\prime)\,dt^\prime
\end{equation}
where $F_{\rm l}(\nu,t)$ is the observed line flux at frequency $\nu$ and
time $t$, $F_{\rm c}(t)$ is the observed ionizing continuum flux at time
$t$, and $\psi(\nu,t)$ is known as the {\it 2-dimensional transfer
  function}.  In writing this equation, it has been assumed that the
ionizing continuum possesses a fixed spectral form and varies only in
normalization. Integrating $\psi(\nu,t)$ over frequency gives the
corresponding 1-dimensional transfer function, $\psi_{\rm 1d}(t)$.  Given
well-sampled data of sufficient quality, this equation can be inverted in
order to obtain the 2-d transfer function.  Much of the attention in BLR
reverberation studies has concentrated upon measuring this transfer
function from large datasets and comparing with theoretical calculations in
order to probe the BLR geometry.

Iron line reverberation mapping is a qualitatively different endeavor.
Unlike the BLR studies, the geometry of the reprocessor (i.e., the
accretion disk) is relatively well understood.  Instead, the unknowns are
the geometry of the X-ray source and the spin of the black hole (i.e., the
geometry of spacetime).  The fact that the X-ray source is very likely
distributed destroys the linear nature of the reverberation problem.  To
see this, suppose that we divide the (distributed) X-ray source into a
number of sub-units, each of which is small enough to be considered as a
point source.  Suppose that the iron line response to the continuum
emission from each subunit is linear in the sense that it obeys equation
(1).  Summing up the contribution from all subunits, the overall observed
line flux can be written as
\begin{equation}
F_{\rm l}(\nu,t)=\int_{-\infty}^{\infty}F_{\rm c}(t-t^\prime)\sum_n
f^{(n)}(t-t^\prime)\psi^{(n)}(\nu,t^\prime)\,dt^\prime
\end{equation}
where $f^{(n)}(t)$ is the fraction of the observed continuum coming from
the $n$th source subunit at time $t$, and $\psi^{(n)}(\nu,t)$ is the 2-d
transfer function for the $n$th subunit.  Since different source subunits
are in spatially different locations, the functions $\psi^{(n)}$ will
generally be different for different $n$.  Thus, the only situation in
which eqn~(2) can be cast into the form of eqn (1) is when all of the
$f^{(n)}$ are time-independent.  This corresponds to the acausal, and hence
unphysical, scenario in which all parts of the distributed X-ray source
vary simultaneously and in direct proportion to one another.  Counting the
number of degrees of freedom, it is clear that equation (2) cannot be
inverted to give $f^{(n)}(t)$ and $\psi^{(n)}(\nu,t)$ in terms of $F_{\rm
  l}(\nu,t)$ and $F_{\rm c}(t)$.

Despite the horrors described above, progress is possible.  We know that
the X-ray light curves of AGN often show extremely rapid flare and
step-like events.  Some of these events occur on times comparable to, or
shorter than, the lightcrossing time of the central black hole (see
discussion in Section~1).  If these events correspond to the activation
of single isolated regions of the X-ray corona, the iron line response to
these events might be describable in reasonably simple terms (i.e., using
linear response theory).  This dictates the approach that we shall take in
this paper.  Rather than addressing the general problem of extracting
information from general forms of $F_{\rm l}(\nu,t)$ and $F_{\rm c}(t)$, we
shall examine the line response in specific scenarios with the aim of
creating specific observational diagnostics.

\section{The disk-flare model}

In this section, we describe the basic model of the accretion disk and
X-ray flare that is utilized throughout the rest of this paper.  Our
assumptions are tailored to the case of Seyfert galaxies since this will be
the first class of systems in which iron line reverberation will be
accessible observationally.  In essence this model is the Kerr geometry
generalization of the model used by RB97.  However, it must be stressed
that whilst RB97 focussed on line emission from within the radius of
marginal stability ($r=r_{\rm ms}$), that is not the principal aim in this
work.  Line emission from within $r=r_{\rm ms}$ {\it is} included in this
model using the simple prescription of RB97 but is only of importance in
certain extreme regimes of parameter space.  Thus, the unmodelled physical
complexities of this region such as the effect of strong shear on radiative
transfer or the possible presence of a dynamically important magnetic
field, do not represent a significant limitation of our model.

\subsection{The accretion disk}

We consider an accretion disk in prograde orbit about a rotating black hole
which has mass $m$ and dimensionless spin-parameter $a$.  The disk is
assumed to be geometrically-thin and is taken to lie in the symmetry plane
of the rotating black hole.  The mass of the disk is assumed to be
negligible compared to the black hole, allowing the space-time of the
region to be described by the Kerr metric.

A crucial physical quantity is the velocity field of the disk material.  We
use the following approximation for the accretion disk velocity field which
is valid due to our assumption of a geometrically-thin disk.  Outside of
the radius of marginal stability ($r>r_{\rm ms}$), we assume that the disk
material lies on circular free-fall orbits.  The small inflow inherent to
the accretion process is not important for our current purposes (i.e.
determining iron line diagnostics) and shall be ignored.  Thus, our model
velocity field is given by $\dot{r}=0$, $\dot{\theta}=0$ together with
equations (A5), (A6), (A8) and (A9).  Within the radius of marginal
stability, such circular orbits are no longer stable.  We assume that
material spirals into the hole on ballistic orbits with the energy and
angular momentum of the disk material at $r=r_{\rm ms}$.  Then, the
velocity field is given by $\dot \theta=0$ together with equations (A5),
(A6), (A7) and the conditions that the energy is $E=E(r_{\rm ms})$ and the
angular momentum is $l=l(r_{\rm ms})$.

To produce appreciable X-ray reflection and iron line fluorescence upon
X-ray illumination, the disk must be optically thick to electron
scattering.  For any reasonable set of AGN parameters, this condition is
easily satisfied outside of the radius of marginal stability (e.g., see
Frank, King \& Raine 1992).  Even within the radius of marginal stability,
mass continuity reveals that the disk is still Thomson thick unless the
accretion rate is rather small (i.e., less that a few per cent of the
Eddington rate; RB97).  In this paper, we shall assume that the disk is
optically-thick at all radii of interest.

\subsection{The X-ray flare}

As motivated by the discussion at the end of Section 2, we will be
considering iron line reverberation due to an isolated X-ray flare.  The
flare is assumed to be a point source, emitting an instantaneous flash,
with Boyer-Lindquist coordinates $(r_{\rm s}, \theta_{\rm s}, \phi_{\rm
  s})$.  The flare is assumed to be an isotropic emitter in the locally
non-rotating frame.

We calculate the illuminating X-ray flux as seen by the disk material,
$F_{\rm X}(r,\phi)$, as follows.  For a particular source position $(r_{\rm
  s}, \theta_{\rm s}, \phi_{\rm s})$ we numerically integrate an isotropic
distribution of photons through the Kerr geometry from the source to the
accretion disk, event horizon, or `infinity' (operationally defined at
$r=1000m$).  We use the expressions of Karas, Vokrouhlick\'y \& Polnarev
(1992), quoted in the Appendix, to enforce this isotropy.  Thus, we can
calculate the illuminating X-ray flux as seen in the inertial frame of the
disk material, taking into account the gravitational focusing effects and
the Doppler effects caused by the orbital motion of the disk material.  To
facilitate our reverberation calculations, we keep track of the time taken
$t_{\rm s}(r, \phi)$ for photons to travel from the source to position $(r,
\phi)$ on the disk.

\subsection{Ionization of the disk material}

The ionization state of the disk has an important bearing on its X-ray
reflection properties and hence must be considered.  Motivated by {\it
  ASCA} observations of Seyfert galaxies (Tanaka et al. 1995, Reynolds
1997, Nandra et al. 1997a), and theoretical studies (Matt, Fabian \& Ross
1993, 1996) we assume that the disk outside of the radius of marginal
stability is `cold' in the sense that it produces a 6.4\,keV fluorescent
iron line upon X-ray illumination\footnote{In higher luminosity AGN, which
  are operating closer to the Eddington limit, even the region outside of
  the radius of marginal stability may be appreciably ionized.  These
  effects have been considered theoretically by Matt, Fabian \& Ross (1993,
  1996), and may have been observed by Nandra et al. (1997b).}.

Inside the radius of marginal stability, the density drops precipitously
and hence photoionization can lead to significant ionization of the disk
material.  We define the X-ray ionization parameter by
\begin{equation}
  \xi(r,\phi)=\frac{4\pi F_{\rm X}(r,\phi)}{n(r)\cos\alpha}, 
\end{equation}
where $F_{\rm X}$ ionizing flux striking the disk (assumed to be a
power-law of photon index $\Gamma=2$ between $13.6\eV$ and $100\keV$),
$n(r)$ is the electron number density, and $\alpha$ is the angle that the
incoming rays make with the disk normal (all quantities measured in the
local inertial frame of the disk material).  The electron density $n(r)$ is
determined from the continuity equation.  It is readily shown that the
expressions of RB97 generalize to the Kerr metric, giving
\begin{equation}
  n=\frac{{\dot m}}{4\pi r h_{\rm disk} (-u^r) m_{\rm p}}, 
\end{equation}
where ${\dot m}$ is the mass accretion rate and $h_{\rm disk}$ is the half
thickness of the disk (see the Appendix for a sketch of the relativistic
derivation of this equation).  We take $h_{\rm disk}/r=10^{-2}$ to be
representative (Muchotrzeb \& Pacy\'nski 1982).

Once the ionization parameter has been determined, the response of the line
emission to the ionization state is computed using a slightly modified form
of the simple prescription of RB97 (based on the detailed calculations of
Matt, Fabian \& Ross 1993, 1996).  In this prescription, each point of the
disk is treated as being in one of four ionization zones.  For
$\xi<100\erg\cm\ps$, we assume a cold iron fluorescence line at $6.4\keV$
with the strength given by the standard neutral slab calculations (e.g.,
George \& Fabian 1991).  For $\xi$ in the range of $100\erg\cm\ps$ to
$500\erg\cm\ps$, we assume no line emission due to the efficient resonance
trapping and Auger destruction of these line photons.  For $\xi$ in the
range $500\erg\cm\ps$ to $5000\erg\cm\ps$, we assume a blend of helium-like
and hydrogen-like iron line emission with a rest-frame energies of
$6.67\keV$ and $6.97\keV$, respectively, with an effective fluorescent
yield for each line the same as that for the neutral case.  For
$\xi>5000\erg\cm\ps$, the material is taken to be completely ionized and no
line emission results.

\subsection{Caveats and realities}

The model described above captures the essentials necessary to perform our
calculations.  However, some of its approximations and simplifications
deserve a brief mention.  One major simplification is that we have chosen
not to use a full transonic disk model, favoring instead the ballistic
approximation made in RB97.  However, this approximation is accurate to
within a few percent due to the assumption that the disk is
geometrically-thin and hence the material orbits {\it are} essentially
ballistic.  This is ample accuracy for the current purposes.

We have also neglected the vertical structure of the accretion disk,
choosing to treat it as possessing a uniform vertical structure.  In the
cold region outside of the radius of marginal stability, this is not an
important omission since the exact ionization balance is unimportant and
the surface layers of gas are optically-thin to the fluorescent iron line
photons.  For most of the relevant regions of parameter space, this
comparatively well understood region dominates the observed iron line
properties.  Within the radius of marginal stability, where ionization
effects are important, the exact vertical structure will effect the
relevant ionization state and hence the iron line emission.  Equally
important will be the complex radiative transfer effects as the reprocessed
X-ray photons (including the fluorescent iron line photons) propagate
through the ionized, strongly shearing, disk material.  Such calculations
are well beyond the scope of this work.  Given that the region inside the
radius of marginal stability usually has a small observable influence, our
simple scheme seems appropriate for application to all foreseeable iron
line observations.

\section{Calculation of observables}

We suppose that the observer is situated at a large distance from the black
hole system and is viewing the system with inclination $i$ (where $i=0$
corresponds to viewing the accretion disk face-on).  We shall define the
Cartesian coordinate system $(x,y)$ on the observer's image plane.  The
origin of this coordinate system is the place where a `radial' ray from the
black hole (i.e., $\phi=\theta={\rm constant}$ in Boyer-Lindquist
coordinates) intercepts the image plane.  Photon paths (i.e., null
geodesics) are numerically integrated through the Kerr metric from points
$(x,y)$ on the image plane (which for practical purposes we place at
$r=1000m$) to the disk plane\footnote{If the condition $r<1.05r_{\rm evt}$
  (where $r_{\rm evt}$ is the radius of the event horizon and is reported
  in the appendix) is satisfied at any point along the photon path, the
  photon is deemed to have entered the event horizon and the integration is
  terminated.}.  The photon paths are integrated using the four constants
of motion: the energy, azimuthal angular momentum, photon rest mass and the
Carter constant.  The relevant equations of motion governing these photon
paths are reported explicitly in the Appendix.  Note that this procedure
assigns a unique point $(r,\phi)$ on the accretion disk to each point
$(x,y)$ on the image plane.

From the results of these integrations, we can evaluate the following
functions which are defined on the image plane.
\begin{enumerate}
\item $R(x,y)$ --- the radius $r$ of the point in the accretion
  disk which is viewed at the point $(x,y)$ in the image plane.
\item $\Phi(x,y)$ --- the azimuthal angle $\phi$ of the point in the
  accretion disk which is viewed at the point $(x,y)$ in the image plane.
\item $g(x,y)$ --- the frequency boost factor of photons that are emitted
  from the accretion disk (with frequency $\nu$ in the inertial frame of
  that point in the accretion disk) and pass through the point $(x,y)$ on
  the image plane (with observed frequency $g(x,y)\nu$).
\item $t(x,y)$ --- the time taken for a photon to be emitted from the
  accretion disk and pass through the point $(x,y)$ on image plane.
\end{enumerate} 
The time dependent line (photon) flux $F_{\rm l}(\nu,t)$ which is driven by
the isolated X-ray flare can then be calculated by
\begin{equation}
  F_{\rm l}(\nu,T)d\nu\,dT\propto\int\int_{T\rightarrow T+dT, \atop
    \nu\rightarrow \nu+d\nu}A(x,y)F_{\rm X}(R, \Phi)g^3\,\,dx\,dy
\end{equation}
where $T(x,y)=t(x,y)+t_{\rm s}\left(R(x,y), \Phi(x,y)\right)$, and the
integration is over those regions of the image plane with $T$ in the range
$T\rightarrow T+dT$ and $\nu=g\nu_0$ (where $\nu_0\equiv 6.4\keV$ is the
rest-frame frequency of the emission line) in the range $\nu\rightarrow
\nu+d\nu$.  $A(x,y)$ is the proper area of the disk element subtended by
the image plane pixel at position $(x,y)$.  This factor accounts correctly
for the general relativistic solid angle transformations.

\section{Reverberation signatures for on-axis flares}

First, we consider iron line reverberation with an X-ray source situated on
the symmetry axis of the black-hole/disk system, at a height $r=h$ above
the disk plane.  Physically, this would be an appropriate geometry if the
X-ray flares occur at high latitude in a geometrically-thick disk-corona.
Alternatively, X-ray flaring from the base of an axial jet would be well
modelled by these calculations.  Figure~1 gives the 2-dimensional transfer
functions for the case of Schwarzschild geometry ($a=0$).  We have assumed
a source height of $h=10Gm/c^2$, a source efficiency of $\eta_{\rm
  x}=0.01$, and have shown four observer inclinations
($i=3\degmark,30\degmark,60\degmark,80\degmark$).  The zero of the time
variable is defined as the moment when the {\it direct} radiation from the
pulsing source reaches the observer.  The 2-d transfer functions are most
readily interpreted by noting that vertical `slices' correspond to line
profiles at some instant in time.

First, consider the almost face-on case (i.e., $i=3\degmark$; note that we
avoid the exactly face-on case so as to avoid the coordinate singularity
present at the poles).  There is an initial delay between the observed
pulse and the response in the line which is simply due to light travel
times.  If one were to image the disk at subsequent times, the line
emitting region would be an expanding ring centered on the disk.
Initially, the line emission will come from the innermost regions of the
disk and will be highly redshifted by gravitational redshifts and the
transverse Doppler effect.  As the line-emitting region expands, these
effects lessen and the observed line frequency tends to the rest-frame
frequency.  

As one considers higher inclination systems, Doppler effects come into play
and the line is broadened.  The time delay between the observed pulse and
the line response is also shortened due to the geometry.  At
moderate-to-high inclinations, a generic feature appears in the line
response whose presence is a direct consequence of relativity.  Soon after
the line profile starts responding to the observed flare, the red wing of
the line fades away.  During these times, the red wing of the line is due
to emission from the {\it front} and {\it receding} portions of the disk,
with gravitational redshifts being the dominant effect.  Some time later,
the observed `echo' of the X-ray flare reaches the {\it back} side of the
disk, whose solid angle at the observer is enhanced by lensing around the
black hole itself.  When the echo reaches this region, the red-wing of the
line dramatically recovers before finally fading away with the rest of the
line response.

An interesting feature can be seen in the high inclination systems (i.e.,
$i=60\degmark$ and $i=80\degmark$).  The transfer functions for these
inclinations show a high-energy `double loop' at $t\sim 20Gm/c^3$
corresponding to distinct high-energy peaks in the iron line profile.  This
is due to iron line emission from ionized helium and hydrogen-like iron
which exists in a narrow annulus just inside the innermost stable orbit.
In time-averaged line profiles, this ionized emission would be seen as
isolated high-energy peaks (e.g., see Fig.~5 of RB97).  Conversely, the
observation of such isolated high-energy peaks in a high inclination source
would be evidence that we are witnessing a sharp change in the ionization
structure of the disk.  This, in turn, would be evidence that we are seeing
the radius of marginal stability around a slowly rotating black hole.

Similarly, Fig.~2 shows the 2-d transfer functions for a maximally rotating
Kerr black hole and the same source parameters.  Again, four inclinations
are shown ($i=3\degmark,30\degmark,60\degmark,80\degmark$).  Note that we
define a maximally rotating hole, following Novikov \& Thorne (1973), to be
one with a spin parameter of $a=0.998$.  The innermost stable orbit, and
hence by assumption the Keplerian part of the accretion disk, is much
smaller than in the Schwarzschild case ($r_{\rm ms}=1.23Gm/c^2$ for
$a=0.998$).  This immediately leads to an interesting phenomenon in the
line reverberation which is most simply illustrated by considering the
face-on case (Fig.~3a).  At a time of $\sim 25Gm/c^3$, the observer sees
{\it two} rings of line emission --- one is propagating outwards into the
disk (as in the Schwarzschild case) and the other is propagating {\it
  inwards} towards the event horizon.  This second ring corresponds to line
photons that have been delayed due to their passage through the strongly
curved space in the near vicinity of the hole (i.e., the Shapiro effect).
This produces a small red bump in the observed line spectrum which moves
progressively to lower energies as time proceeds.  In the case of a
truly-maximal Kerr hole (i.e., one with $a=1$), our model would have the
Keplerian disk extending exactly to the horizon and this inward moving ring
would be seen to tend asymptotically to the event horizon (with ever
increasing redshifts).  This phenomenon is further illustrated in Fig.~3,
where we show the observed line profiles at various times.  This effect is
more prominent for sources closer to the disk (i.e., smaller $h$).  Our
calculations show these redwards moving bumps to be generic features of
line reverberation around near-extremal kerr holes, and hence may be
considered direct observational signatures of near-extremal kerr geometry.


In the immediate future (i.e., prior to the launch of {\it XMM}), only the
time-lag of the pulse and response is within observational reach.  It is
interesting to compare the minimum time-lags derived from our fully
relativistic calculations with those expected in Euclidean geometry.  In
Euclidean geometry, it is readily shown that the minimum time lag (i.e., the
time between the observed continuum pulse and the {\it first} observed
response from the disk line) is
\begin{equation}
  t_{\rm delay}=\frac{2h}{c}\cos i.
\end{equation}
This is plotted in Fig.~4 and compared to the results from our relativistic
calculations.   Note that for low inclinations, there is an extra time-lag
due to the time taken for photons to propagate through the strongly curved
space in the vicinity of the black hole.   Noting that $i$ can be obtained
from time averaged line profiles, the observation of such a lag allows $h$
to be determined.   

Further in the future, high throughput instruments such as {\it XMM} and
the {\it Constellation X-ray Mission} (formerly {\it HTXS}) will be able to
measure the time response of several different bands covering the line
profile.  Measuring the response of the line profile to a dramatic event in
the continuum light curve will allow the 2-d transfer function (for that
event; see Section 2) to be mapped out.  Again, noting that $i$ can be
constrained from the time-averaged line profiles, comparing such data to
theoretical transfer functions will allow the mass and spin of the hole to
be constrained.  In a future publication, we will present detailed
simulations of {\it Constellation-X} observations and determine the
possible constraints that can be applied to the source parameters.

\section{Reverberation signatures for off-axis flares}


If the X-ray emission of accreting black holes does indeed originate in an
accretion disk corona, we would expect most of the X-ray flares to occur
significantly away from the symmetry axis.  This clearly expands the
parameter space that we must explore: the instantaneous motion of the flare
as well as its latitude and azimuth must be specified in order to calculate
the corresponding illumination pattern on the disk.  A detailed
investigation of iron line reverberation from off-axis flares will form the
basis for a future publication.  However, for completeness we present a
representative off-axis calculation to highlight the principal effects.  In
these preliminary calculations, the flare is assumed to be an isotropic
emitter in the locally non-rotating frame.

In Fig.~5 we show the 2-d transfer functions for an off-axis flare with
$r_{\rm s}=10r_{\rm g}$ and $\theta_{\rm s}=70\degmark$ on both the
receding side of the disk ($\phi_{\rm s}=90\degmark$; panel a) and the
approaching side of the disk ($\phi_{\rm s}=270\degmark $; panel b).  We
assume an extreme-Kerr hole ($a=0.998$) and an observer inclination of
$i=30\degmark$ (appropriate for a typical Seyfert 1 nucleus).  These
figures are to be contrasted with Fig.~3b, which shows the on-axis flare
case for the same spin-parameter, inclination and value of $r_{\rm s}$.

As expected, X-ray flares close to the disk tend to produce briefer and
narrower line responses.  In the limiting case of flares in a a thin,
disk-hugging X-ray emitting corona (i.e., the limit $\theta\rightarrow
90\degmark$ of the above calculations) the line response to the continuum
flare will be essentially immediate and monochromatic. In other words, the
2-d transfer function becomes a delta-function
$\psi(\nu,t)=\phi_0\delta(\nu-\nu_f)\delta(t)$ where the observed frequency
of the line response $\nu_f$ is dependent on the radius and azimuth of the
flare, the spin-parameter of the hole, and the observer inclination.

\section{MCG$-$6-30-15 revisited}


As an illustration of the use of iron line reverberation, we return to
the case of MCG$-$6-30-15.  As mentioned in the introduction, the
very-broad state of the iron line in this Seyfert 1 galaxy as found by
Iwasawa et al. (1996) can be reasonably modeled as either emission from a
disk around a near-extremal Kerr hole (Iwasawa et al. 1996; Dabrowski et al.
1997), or emission from within the innermost stable orbit of a slowly
rotating hole (RB97).  A basic degeneracy exists insofar as both of these
models can be made consistent with current time-averaged iron line data.
Reverberation studies are the key to unambiguously distinguishing these
effects.

A key ingredient of the RB97 model is an X-ray source which is displaced
from the disk by heights of $h\sim 4Gm/c^2$ to $h\sim 12Gm/c^2$ (depending
on exactly which state the system is in).  Physically, this was imagined to
be X-ray emission from near the base of a jet or an extended corona.  The
region inside the innermost stable orbit is then strongly illuminated by
this emission due to gravitational focusing and blueshifting as the photons
`fall' onto the disk.  Additionally, in order for this region not to be
almost completely ionized by this flux, the X-ray efficiency of the source
has to be rather low ($\eta_{\rm x}\sim 10^{-3}$).  Figure~6a shows the 2-d
transfer function of the RB97 scenario for the very-broad state of the iron
line in MCG$-$6-30-15 ($i=27\degmark$, $a=0$, $h=4Gm/c^2$ and $\eta_{\rm
  x}=10^{-3}$).

On the other hand, the extremal-Kerr model for this line (Dabrowski et al.
1996) assumes a thin-coronal geometry.  In this case, a flare from the
continuum source will induce an immediate monochromatic response in the
line as described in Section 6.  This should be easily distinguishable from
the scenario of the previous paragraph.  Of course, a possible hybrid model
for the very broad line is a displaced source above a disk around a
rapidly-rotating black hole (Martocchia \& Matt 1996).  The transfer
function for this case is shown in Fig.~6b.  This will be rather more
difficult to distinguish from the RB97 model.  However, sufficiently
detailed constraints on the 2-d transfer function, especially on the
re-emergence of the red wing of the line, can distinguish these models.

A crucial parameter in these studies is the light crossing time of the
gravitational radius ($t_{\rm g}=Gm/c^3$).  For the case of MCG$-$6-30-15,
the multiwaveband analysis of Reynolds et al. (1997) estimated a
lower-limit on the bolometric luminosity of the AGN to be $L\approxgt
1\times 10^{44}\ergps$.  The fact that the disk (outside of the innermost
stable orbit) is in a physical state capable of producing cold iron
fluorescence suggests that the system is operating at no more that 10 per
cent of its Eddington rate.  Thus, a lower-limit to the mass of the black
hole in MCG$-$6-30-15 is $M\approxgt 10^7\Msun$.  The corresponding
timescale is $t_{\rm g}\approxgt 50\s$.  Given the nature of these limits,
this timescale could easily be an order of magnitude larger.  Taking the
conservative figure of $t_{\rm g}=50\s$, the RB97 model for the
`very-broad' state of the system predicts a time lag of approximately
$400\s$ between a continuum flare and the response of the iron line.  

\section{Conclusions}

We have examined how broad X-ray iron lines respond to variability in the
primary X-ray source, i.e., we have addressed the reverberation of
fluorescent lines from the innermost regions of an accretion disk around a
black hole.

In any physical model for the X-ray source (e.g., disk-corona or jet), the
source is likely to be extended with a size comparable to the region of the
disk producing the line.  In this case, there is no well-defined transfer
function that relates the continuum light curve to the time-dependent line
profile.  Hence, it is not possible to perform a general inversion and
constrain the source geometry or black hole spin-parameter from
continuum/line light curves.  The analogous problem in optical/UV BLR
studies {\it is} invertible since the exciting UV source is very much
smaller than the BLR.  Despite this fundamental non-invertibility,
progress is possible since AGN light curves are known to undergo dramatic
events which presumably correspond to the activation of a new localized
active region.  The response of the iron line emission to the activation of
such a region will possess clean reverberation signatures (and can be
described in terms of a transfer function {\it for that particular active
  region}).

Our detailed calculations assume an X-ray source which is located on the
symmetry axis of the black-hole/disk system and at a height $h$ above the
disk plane.  We have computed the line flux as a function of energy and
time when the X-ray source emits a $\delta$-function pulse of hard X-ray
radiation.  This amounts to computing the 2-dimensional transfer function
for the problem.  The most basic consequence of having the X-ray source
displaced from the disk is a time lag between the observed continuum pulse
and the first response of the line.  This lag is slightly longer than, but
approximated by, the Euclidean result presented in equation (6).  The
`excess' lag is due to the passage of the photons through the strongly
curved spacetime in the vicinity of the black hole. The detection of this
lag (which should be within reach of {\it RXTE}) will allow the height of
the source, $h$, to be measured.   We also briefly address the off-axis
flare case.

From our computations of the general 2-d transfer functions, we note several
interesting features.  
\begin{enumerate}
\item A generic feature of line variablity in moderate-to-high inclination
  systems is a `re-emergence' of the red-wing some time after the observed
  X-ray flare.  This is gravitationally redshifted emission from the
  back-side of the disk that has been enhanced by gravitational lensing
  around the hole.  The exact nature and timing of this `re-emergence'
  effect depends upon the mass and spin of the black hole, and should be
  fairly readily observable by future high-throughput spectrometers.
\item An observer viewing line reverberation from a disk around a rapidly
  rotating hole will see a distinct bump in the line profile that
  progresses to lower energies as time proceeds.  If the observer could
  image the disk, this spectral feature would originate from a ring of
  emission moving asymptotically towards the horizon.  It corresponds to
  photons that have been delayed by their progress through the strongly
  curved spacetime around the black hole.  Since the formation of such a
  feature requires a relatively cold disk close to the event horizon, its
  detection would indicate the existence of a rapidly rotating black hole.
  Indeed, our calculations show this feature to be a good signature of near
  extremal kerr geometry.  We note, however, that a detection of this
  feature would be challenging since it is weak and would be largely
  swamped by the primary continuum emission.
\item A high inclination observer viewing a line from around a slowly
  rotating hole may observe a distinct high-energy peak on the iron line
  profile.  This corresponds to line emission from ionized material within
  the radius of marginal stability.  Although the signature of this
  emission is more clearly separated from the bulk of the line emission if
  one considers time-resolved line profiles, this feature is also
  observable in the time-averaged line profile.  The detection of this
  distinct high-energy peak, which should be easily within reach of {\it
    XMM} and {\it Constellation}, would strongly argue for a Schwarzschild
  black hole.
\end{enumerate}

Starting with {\it Constellation}, future instruments should be able to
measure the detailed response of the iron line profile to rapid changes in
the primary X-ray continuum.  Given such data, it will be possible to place
unprecedented constraints on the geometry of the system, the mass of the
black hole, and even the black hole's spin.

\section*{Acknowledgments}

We thank Mike Nowak and J\"orn Wilms for many useful discussions.  This
work has been supported by the National Science Foundation under grant
AST-9529175 (CSR, MCB), NASA under grant NAG-6337 (CSR, MCB), PPARC (AY),
and the Royal Society (ACF).


\appendix

\section{Some properties of the Kerr metric}

In this appendix, we collect together some standard results on the
properties of the Kerr metric which we use in the main body of the paper.
These results can be found in several standard texts, but we mostly shall
follow the techniques and notation of Shapiro \& Teukolsky (1983).

\subsection{The metric and the orbital motion of accretion disk material}

In standard Boyer-Lindquist coordinates, the Kerr metric is
\begin{equation}
  ds^2=-\left(1-\frac{2mr}{\Sigma}\right)dt^2 -
  \frac{4amr\sin^2\theta}{\Sigma}dt\,d\phi +
  \frac{\Sigma}{\Delta}dr^2+\Sigma\,d\theta^2+\left(r^2+a^2+\frac{2a^2mr\sin^2\theta}{\Sigma}\right)\sin^2\theta\,d\phi^2
\end{equation}
where
\begin{eqnarray}
  \Delta &=& r^2-2mr+a^2\\
  \Sigma &=& r^2+a^2\cos^2\theta
\end{eqnarray}
and we have chosen units such as to set $G=c=1$.  The event horizon of the
black hole is given by the outer root of $\Delta=0$, i.e. $r_{\rm
  evt}=m(1+\sqrt{1-a^2})$.

By assumption, the accretion disk lies in the equatorial plane of the
rotating black hole and orbits in the prograde sense.  Since the disk is
assumed to be geometrically-thin, pressure forces are negligible and the
disk material will follow free-fall paths through the metric.  From the
metric, we form the (restricted) Lagrangian ${\cal L}_{\rm d}$ describing
free-fall particle orbits in the equatorial plane,
\begin{eqnarray}
  {\cal
  L}_{\rm d}&=&\frac{1}{2}\left(\frac{ds}{d\tau}\right)^2_{\theta=\pi/2}\\\nonumber
&=&\frac{1}{2}\left[-\left(1-\frac{2m}{r}\right){\dot t}^2
  -\frac{4am}{r}{\dot t}{\dot \phi}+\frac{r^2}{\Delta}{\dot r}^2 +
  \left(r^2+a^2+\frac{2ma^2}{r}\right){\dot \phi}^2\right]
\end{eqnarray}
where $\tau$ is an affine parameter that can (and will) be taken to be
proper time in the case of massive particles, and the dot denotes
differentiation with respect to $\tau$.  The Euler-Lagrange equations
together with the fact that the Kerr metric is stationary and axisymetric
(i.e. $ds^2$ has no explicit dependence upon $t$ or $\phi$) imply that for a
free-fall path
\begin{eqnarray}
  {\dot t}&=&\frac{(r^3+a^2r+2a^2m)E-2aml}{r\Delta}\\
  {\dot \phi}&=&\frac{(r-2m)l+2amE}{r\Delta}
\end{eqnarray}
where $E$ is the conserved energy ($E=-\partial {\cal L}_{\rm d}/\partial
{\dot t}$), and $l$ is the conserved aximuthal angular momentum
($l=\partial {\cal L}_{\rm d}/\partial {\dot \phi}$).  For massive
particles, the conservation of rest-mass gives ${\cal L}_{\rm
  d}=-\frac{1}{2}$ which leads to the following radial equation of motion:
\begin{equation}
  r^3{\dot r}^2 = E^2(r^3+a^2r+2ma^2)-4amEl-(r-2m)l^2-rm^2\Delta.
\end{equation}
Equations~(A5)--(A7) are the basic equations of motion for a massive particle in
the equatorial plane of the Kerr metric.

Demanding that the orbits be circular (i.e. ${\dot r}={\ddot r}=0$)
determines the energy and angular momentum as a function of radius.  The
result is
\begin{eqnarray}
  \frac{E(r)}{m}&=&\frac{r^2-2mr+a\sqrt{mr}}{r(r^2-3mr+2a\sqrt{mr})^{1/2}}\\
  \frac{l(r)}{m}&=&\frac{\sqrt{mr}(r^2-2a\sqrt{mr}+a^2)}{r(r^2-3mr+2a\sqrt{mr})^{1/2}}
\end{eqnarray}
A very important result is that such circular orbits are only stable for
$r>r_{\rm ms}$, where the critical radius $r_{\rm ms}$ is called the {\it
  radius of marginal stability} and is given by
\begin{equation}
  r_{\rm ms}=m\left(3+Z_2-\left[(3-Z_1)(3+Z_1+2Z_2)\right]^{1/2}\right)
\end{equation}
where
\begin{equation}
  Z_1=1+\left(1-\frac{a^2}{m^2}\right)\left[\left(1+\frac{a}{m}\right)^{1/3}+\left(1-\frac{a}{m}\right)^{1/3}\right]
\end{equation}
\begin{equation}
  Z_2=\left(3\frac{a^2}{m^2}+Z_1^2\right)^{1/2}
\end{equation}
This marginally stable orbit is also sometimes referred to as the {\it
  innermost stable orbit}.  Hence $r_{\rm ms}=6m$ for a Schwarzschild hole
($a=0$).  For prograde orbits around a Kerr black hole, $r_{\rm ms}<6m$ and
tends to $r_{\rm ms}\rightarrow m$ as $a\rightarrow 1$.  Note that $r_{\rm
  evt}\rightarrow m$ also as $a\rightarrow 1$. In other words, in the limit
of a maximally rotating black hole, the radius of marginal stability
extends all of the way down to the event horizon.

\subsection{Continuity equation and the density of the disk}

The fully-relativistic (baryon number) continuity equation is
\begin{equation}
  \left(\rho u^\mu\right)_{;\mu}=0,
\end{equation}
which, using the metric connection, can be expressed in terms of the
determinant of the metric tensor (${\rm det}\,g$) as
\begin{equation}
  \left(\rho u^\mu\sqrt{-{\rm det}\,g}\right)_{,\mu}=0.
\end{equation}
By writing the metric tensor in matrix form, it is readily shown that for
flows in the equatorial plane ${\rm det}\,g=r^4$, and so equation~(A14)  can
be integrated to give
\begin{equation}
  2\pi r\sigma(r) u^r={\dot m}
\end{equation}
where ${\dot m}$ is the mass accretion rate and $\sigma (r)$ is the surface
density.  Assuming the disk to possess a uniform density structure in the
vertical direction and have a half height $h_{\rm disk}$, the density of
the disk at a given radius is
\begin{eqnarray}
  n&=&\frac{\sigma}{2h_{\rm disk}m_{\rm p}}\\\nonumber
&=&\frac{{\dot m}}{4\pi r h_{\rm disk} (-u^r)m_{\rm p}}.
\end{eqnarray}
This is equation~(2) of the main text.

\subsection{Photon propagation in the Kerr metric}

Generalizing the Lagrangian to include orbits not in the plane, we get
\begin{equation}
    2{\cal L}=-\left(1-\frac{2mr}{\Sigma}\right){\dot t}^2 -
  \frac{4amr\sin^2\theta}{\Sigma}{\dot t}\,{\dot \phi} +
  \frac{\Sigma}{\Delta}{\dot
  r}^2+\Sigma\,{\dot \theta}^2+\left(r^2+a^2+\frac{2a^2mr\sin^2\theta}{\Sigma}\right)\sin^2\theta\,{\dot \phi}^2
\end{equation}
The Euler-Lagrange equations, together with stationarity and axisymmetry
give
\begin{eqnarray}
  {\dot t}&=&\frac{E\left[\Sigma(r^2+a^2)+2mra^2\sin^2\theta
    \right]-2amrl}{(\Sigma-2mr)(r^2+a^2)+2mra^2\sin^2\theta}\\
  {\dot
    \phi}&=&\frac{2amrE\sin^2\theta+(\Sigma-2mr)l}{(\Sigma-2mr)(r^2+a^2)\sin^2\theta+2mra^2\sin^4\theta}
\end{eqnarray}
The conservation of photon rest mass gives ${\cal L}=0$ which leads to the
radial equation of motion
\begin{equation}
  {\dot r}^2=\frac{\Delta}{\Sigma}({\dot t}E-{\dot \phi}l-{\dot \theta}^2\Sigma)
\end{equation}
The $\theta$-equation of motion is most readily obtained from the Carter
constant
\begin{equation}
  Q=p_\theta^2-a^2E\cos^2\theta +l^2\cot^2\theta
\end{equation}
where $p_\theta$ is the canonical $\theta$-momentum given by
\begin{equation}
  p_\theta=\frac{\partial {\cal L}}{\partial {\dot \theta}}=\Sigma{\dot \theta}
\end{equation}
Hence,
\begin{equation}
  {\dot \theta}^2=\frac{Q+a^2E\cos^2\theta-l^2\cot^2\theta}{\Sigma^2}
\end{equation}
Equations~(A18)--(A20) and (A23) are differential equations of motion for a
photon in the Kerr metric.  Note that since the equations of motion only
determine $\dot{r}^2$ and $\dot{\theta}^2$, the initial signs of $\dot{r}$
and $\dot{\theta}$ must be explicitly given.  Further, the numerical
integrator must search for turning points in $r$ and $\theta$ along the
path and explicitly change the sign of $\dot{r}$ and $\dot{\theta}$ at such
turning points.

\subsection{Enforcing isotropy in the X-ray source frame}

We assume the X-ray source to be in a locally non-rotating frame at
$(r_{\rm s}, \theta_{\rm s}, \phi_{\rm s})$.  Following Karas,
Vokrouhlick\'y \& Polnarev (1992), we define $\alpha_{\rm s}$ and
$\beta_{\rm s}$ to be the polar and aximuthal angles on the sky as seen by
an observer situated at the X-ray source (see Fig.~1 of Karas,
Vokrouhlick\'y \& Polnarev).  Consider a photon emitted from this source.
The constants of motion $l$ and $Q$ characterizing its subsequent path are
given by
\begin{eqnarray}
l&=&\left(\frac{A^{1/2}\sin\theta\sin\alpha_{\rm s}\sin\beta_{\rm
      s}}{\Sigma\epsilon} \right)_{r=r_{\rm s}, \theta=\theta_{\rm s}},\\
Q&=&\left(\frac{(r^2+a^2-al)^2}{\Delta}-\frac{\Sigma\cos^2\alpha}{\epsilon^2}-l^2+2al-a^2\right)_{r=r_{\rm
      s}, \theta=\theta_{\rm s}},
\end{eqnarray}
where
\begin{eqnarray}
A&=&(r^2+a^2)^2-\Delta a^2\sin^2\theta,\\
\epsilon &=&
A^{1/2}\left(\Sigma^{1/2}\Delta^{1/2}+2ar\Sigma^{-1/2}\sin\theta\sin\alpha_{\rm
    s}\sin\beta_{\rm s} \right).
\end{eqnarray}
To enable the integration of a given photon away from the source, it
remains to determine the initial signs of $\dot{r}$ and $\dot{\theta}$.
Substituting eqn (A24)-(A27) into (A18)-(A23), we can examine $\dot{r}^2$
and $\dot{\theta}^2$ as a function of $\alpha_s$ and $\beta_s$.  The values
of $\alpha_s$ and $\beta_2$ where $\dot{r}^2=0$ (or $\dot{\theta}^2=0$)
delineate the regions of the source's local sky in which $\dot{r}$ (or
$\dot{\theta}$) is initially positive or negative.  In fact, the assumption
of a locally non-rotating source leads to the Euclidean-like result:
\begin{eqnarray}
\dot{r}>0&\hbox{ for }&0\le\alpha_{\rm s}<\pi/2,\\
\dot{r}<0&\hbox{ for }&\pi/2<\alpha_{\rm s}\le\pi,\\
\dot{\theta}>0&\hbox{ for }&|\beta_{\rm s}|<\pi/2,\\
\dot{\theta}<0&\hbox{ for }&|\beta_{\rm s}|>\pi/2.
\end{eqnarray}


\begin{thebibliography}{}
\bibitem[]{} Abramowicz M.~A., Kato S., 1989, ApJ, 336, 304
\bibitem[]{} Campana S., Stella L., 1993, MNRAS, 264, 395
\bibitem[]{} Campana S., Stella L., 1995, MNRAS, 272, 585
\bibitem[]{} Chen X., Taam R.~E., 1993, ApJ, 412, 254
\bibitem[]{} Dabrowski Y., Fabian A.~C., Iwasawa K., Lasenby A.~N., Reynolds
  C.~S., 1997, 288, L11
\bibitem[]{} Fabian A.~C., Rees M.~J., Stellar L., White N.~E., 1989, MNRAS,
  238, 729
\bibitem[]{} Fabian A.~C. et al. 1995, MNRAS, 277, L11
\bibitem[]{} George I.~M., Fabian A.~C., 1991, MNRAS, 249, 352
\bibitem[]{} Iwasawa K. et al., 1996, MNRAS, 282, 1038
\bibitem[]{} Karas V., Vokrouhlick\'y D., Polnarev A.~G., 1992, MNRAS,
259, 569
\bibitem[]{} Laor A., 1991, ApJ, 376, 90
\bibitem[]{} Martocchia A., Matt G., 1996, MNRAS, 282, L53
\bibitem[]{} Matt G., Fabian A.~C., Ross R.~R., 1993, MNRAS, 262, 179
\bibitem[]{} Matt G., Fabian A.~C., Ross R.~R., 1996, MNRAS, 278, 1111
\bibitem[]{} Matt G., Perola G.~C., 1992, MNRAS, 259, 433
\bibitem[]{} Matt G., Perola G.~C., Piro L., 1991, A\&A, 247, 25
\bibitem[]{} Muchotrzeb B., Pacy\'nski B., 1982, Acta Astr., 32, 1
\bibitem[]{} Nandra K., George I.~M., Mushotzky R.~F., Turner T.~J., Yaqoob
  T., 1997a, ApJ, 477, 602
\bibitem[]{} Nandra K., George I.~M., Mushotzky R.~F., Turner T.~J., Yaqoob
  T., 1997b, ApJ, 488, 91
\bibitem[]{} Novikov I.~D.,  Thorne K.~S., 1973, in {\it Black Holes}, eds
  C.~DeWitte \& B.~S.~DeWitte (Gordon and Breach Science Publicaters, New
  York), P.344
\bibitem[]{} Page D.~N., Thorne K.~S., 1974, ApJ, 191, 499
\bibitem[]{} Peterson B.~M., 1993, PASP, 105, 247
\bibitem[]{} Reynolds C.~S., Begelman M.~C., 1997, ApJ, 488, 109 (RB97)
\bibitem[]{} Reynolds C.~S., Fabian A.~C., Nandra K., Inoue H., Kunieda H.,
  Iwasawa K., 1995, MNRAS, 277, 901
\bibitem[]{} Reynolds C.~S., Ward M.~J., Fabian A.~C., Celotti A., 1997,
  MNRAS, 291, 403
\bibitem[]{} Stella L., 1990, Nat, 344, 747
\bibitem[]{} Tanaka Y. et al.,  1995, Nat, 375, 659
\bibitem[]{} Young A., Fabian A.~C., Ross R.~R., 1998, MNRAS, submitted
\end{thebibliography}
\end{document}